\documentclass[showkeys, twocolumn]{revtex4}
\usepackage{graphicx}
\usepackage[english]{babel}
\usepackage[utf8]{inputenc}
\usepackage{subfig}
\usepackage{floatrow}
\usepackage{comment}
\usepackage{amsmath}

\begin{document}

\title{Stability of Highly Hydrogenated Monolayer Graphene in Ultra-High Vacuum and in Air}

\author{Alice Apponi\footnote{Corresponding author. E-mail: alice.apponi@roma3.infn.it}$^{1,2}$}
\author{Orlando Castellano$^{1,2}$}
\author{Daniele Paoloni$^{1,2}$}
\author{Domenica Convertino$^3$} 
\author{Neeraj Mishra$^{3,4}$}
\author{Camilla Coletti$^{3,4}$} 
\author{Andrea Casale$^{5}$}
\author{Luca Cecchini$^{8}$}
\author{Alfredo G. Cocco$^{6}$}
\author{Benedetta Corcione$^{7,8}$}
\author{Nicola D'Ambrosio$^{6}$}
\author{Angelo Esposito$^{7,8}$}
\author{Alfredo Ferella$^{6,9}$}
\author{Marcello Messina$^{6}$}
\author{Francesco Pandolfi$^{8}$}
\author{Francesca Pofi$^{6,10}$}
\author{Ilaria Rago$^{8}$}
\author{Nicola Rossi$^{6}$}
\author{Sammar Tayyab$^{7,8}$}
\author{Ravi Prakash Yadav$^{7,8}$}
\author{Federico Virzi$^{6,9}$}
\author{Carlo Mariani$^{7,8}$}
\author{Gianluca Cavoto$^{7,8}$}
\author{Alessandro Ruocco$^{1,2}$}

\affiliation{$^1$Dipartimento di Scienze, Universit\`a degli Studi di Roma Tre, Via della Vasca Navale 84, 00146 Roma, Italy}
\affiliation{$^2$INFN Sezione di Roma Tre, Via della Vasca Navale 84, 00146 Roma, Italy}
\affiliation{$^3$Center for Nanotechnology Innovation @NEST, Istituto Italiano di Tecnologia, Pisa, Italy}
\affiliation{$^4$Graphene Labs, Istituto italiano di tecnologia, Via Morego 30, I-16163 Genova, Italy}
\affiliation{$^5$Department of Physics, Columbia University, 538 West 120th Street, New York, New York 10027, USA}
\affiliation{$^6$ INFN-LNGS, Via G. Acitelli 22, 67100 Assergi-L'Aquila, L'Aquila, Italy}
\affiliation{$^7$Sapienza Universit\`a di Roma, Piazzale Aldo Moro 5, 00185 Roma, Italy}
\affiliation{$^8$INFN Sezione di Roma, Piazzale Aldo Moro 5, 00185 Roma, Italy}
\affiliation{$^{9}$Department of Physics and Chemistry, University of L'Aquila, 67100 L'Aquila, Italy}
\affiliation{$^{10}$Gran Sasso Science Institute, Viale Francesco Crispi 7, 67100 L'Aquila, Italy}

\begin{abstract}
The stability of hydrogenated monolayer graphene was investigated via X-ray photoemission spectroscopy (XPS) for two different environmental conditions: ultra-high vacuum (UHV) and ambient pressure. The study is carried out by measuring the C 1s line shape evolution for two hydrogenated samples one kept in the UHV chamber and the other progressively exposed to air. In particular, the $sp^3$ relative intensity in the C 1s core-level spectrum, represented by the area ratio $\frac{sp^3}{sp^2+sp^3}$, was used as a marker for the hydrogenation-level. After four months in UHV, it resulted almost unchanged within the experimental uncertainty. Thus, a long-term stability of hydrogenated monolayer graphene was found, that indicates this material as a good candidate for hydrogen (or tritium) storage as long as it is kept in vacuum. On the other hand, the C 1s spectrum of the sample exposed to air shows a significant oxidation. A rapid growth up to saturation of the carbon oxides was observed with a time constant $\tau$ = 2.8 $\pm$ 1.2 hours. Finally, the re-exposure of the oxidised sample to atomic hydrogen was found to be an effective method for the recovery of hydrogenated graphene.  The CH stretching mode was measured via electron energy loss spectroscopy as direct footprint of hydrogenated graphene recovery.
\end{abstract}

\keywords{Graphene, Hydrogenated graphene, Hydrogen stability, Hydrogen storage, X-ray photoemission spectroscopy, Electron energy loss spectroscopy.}

\maketitle

\section{Introduction}
Ever since its first isolation from graphite \cite{Novoselov1}, graphene has captured great attention from the scientific community. Graphene is a 2D material of $sp^2$-coordinated carbon atoms arranged in a honeycomb lattice. The unique electrical, mechanical and thermal properties of this material make it attractive for numerous fields of application in science and technology  \cite{Novoselov, PAPAGEORGIOU}. Moreover, the possibility to chemically functionalise the graphene lattice with hydrogen atoms is a further promising feature. When hydrogen atoms attach to the $sp^2$-coordinated carbon atoms of graphene, a distortion of the lattice toward an $sp^3$ configuration leads to the transition from a zero-gap (graphene) to a wide-gap semiconductor (fully hydrogenated graphene) \cite{Boukhvalov, Sofo}. Due to these unique properties of broad interest, the hydrogenation of graphene has been widely investigated in the last decades both in terms of graphene synthesis and structure and in terms of hydrogenation techniques \cite{Luo, Burgess, Zhao, Elias, Ryu, Luo2, Felten, Balog, Panahi, Paris, Haberer, NPG21, NPG22, HYDRO}. 

In the context of sustainable and alternative source of energy, the use of hydrogen as a fuel is one of the most promising and studied possibilities. The common solutions for hydrogen storage are indeed the cryogenic liquid form or the compressed gas state, both presenting considerable disadvantages first of which a safety related issue \cite{Dillon, Kula}. 
Along with metal hydrides, a  valuable alternative is represented by carbon-based structures, which allow to have a safe, reversible, compact and high-loading - 1:1 proportion of H:C eventually - solid state form hydrogen storage \cite{Dillon, Kula, Alekseeva}. 

Tritium is a radioactive hydrogen isotope that has the same chemical properties of its stable counterpart. Great interest in tritium handling lies in its use as a sustainable fuel for next-generation nuclear fusion energy reactors \cite{Treview}. It follows that, as hydrogen can be stored in a solid state form chemically bonded to graphene, the analogous can be achieved with tritium \cite{Zeller}. Therefore, graphene can represent also a promising alternative for tritium storage to the most widely adopted uranium-alloys \cite{Treview}. Furthermore, graphene plays a major role in the tritium handling chain as it can be employed for hydrogen isotopes separation via electrochemical pumping, thus exploiting the graphene selective permeation \cite{Geim, Yasuda}. 

The bonding of atomic tritium to graphene is also important in neutrino physics, for example in the PTOLEMY project \cite{Cocco, PTOLEMY4, Apponi_2022}. This experiment aims to measure the neutrino mass with an unprecedented energy resolution and, eventually, unveil the cosmic neutrino background from the study of the endpoint of the $\beta$-spectrum of tritium. One of the most important novelties introduced by the PTOLEMY experiment is a solid state target, with atomic tritium bonded to graphene nanostructures (\emph{i.e.} nanoporous graphene, stacked monolayer graphene sheets or carbon nanotubes). 

In all the above mentioned frameworks, both involving hydrogen and tritium, the stability of the functionalised graphene surface in different environments is a critical topic - especially to avoid ambient contamination with tritium. The thermal desorption of hydrogen from graphene in different atmospheres (N$_2$, Ar, H$_2$) or in vacuum was investigated experimentally and theoretically for temperatures up to 900 $^\circ$C \cite{Delfino, Fournier, Kula_2016, Whitener}. On the other hand, the dehydrogenation and degradation of hydrogenated graphene at constant room temperature both in air and in different chemical (oxidant) environments, has been studied in a few works via Raman spectroscopy, infrared spectroscopy and resistance experiments \cite{Kula_2016, Whitener, Kaczmarek}. However, a systematic investigation of the H-C chemical and bonding stability for storage purposes, also in view of the application to tritium, is still lacking.

In this work, an extensive study on the stability of hydrogenated monolayer graphene in ultra-high vacuum and in air via X-ray photoemission spectroscopy (XPS) is reported. The recovery of the hydrogenated graphene after oxidation in air with atomic hydrogen exposure is also investigated via both XPS and electron energy loss spectroscopy (EELS). 
The effect of the hydrogen bonding to graphene is to distort the $sp^2$ coordination of carbon atoms toward a $sp^3$-like configuration, with the hydrogen \emph{pulling out} one carbon atom from the lattice plane \cite{Boukhvalov, Sofo, HYDRO}. Therefore, with XPS it is possible to study the C 1s core-level spectrum looking at the $sp^3$ contribution as a marker of graphene hydrogenation and eventually investigate the presence of carbon oxides. On the other hand, EELS measurements allow to observe the CH stretching mode as a direct footprint of the hydrogen bonding to carbon \cite{DiFilippo}.

Finally, in the case of tritium bonded to graphene, one should take into account also the possible degradation of the surface due to the radioactivity of the isotope. Therefore, some considerations on the stability of tritiated graphene are discussed on the basis of the experimental experience with electron and ion bombardment of surfaces.

\section{Sample preparation and experimental methods}
\label{sec:prep}

The samples investigated in this work are prepared at CNI@NEST laboratory in Pisa following the procedure thoroughly described in \cite{Apponi2024}. They consist in polycrystalline monolayer graphene grown via chemical vapor deposition (CVD) on electropolished copper \cite{Miseikis_2015, Convertino2020}, then transferred onto a transmission electron microscopy (TEM) grid with the standard wet etching technique \cite{Miseikis_2015, Li2009, D1CP04316A}. The grids employed for graphene transfer are commercial TEM grids made of nickel (Ted Pella Inc. G2000HAN) without any additional mesh or film on top. After preparation, the samples are inserted in the ultra-high vacuum (UHV) chamber of the LASEC laboratory in Roma Tre University, where a 550 $^\circ$C annealing is performed in order to clean the graphene from PMMA (polymethyl methacrylate) residues due to the transfer procedure \cite{Apponi2024}. The hydrogenation is then carried out by exposing the samples to thermally-cracked atomic hydrogen (FOCUS EFM-H) while keeping the hydrogen partial pressure at 3.6 $\cdot 10^{-6}$ mbar. The hydrogenated samples herein investigated are the same described in \cite{HYDRO}, where a first sample (sample A) reached the hydrogen saturation level (61$\%$ $sp^3$) after a dose of 320 kL (where 1 L = 1.33 $\cdot 10^{-6}$ mbar $\times$ 1 s), while the second one (sample B) after 260 kL (100$\%$ $sp^3$). Therefore, the initial status of the samples investigated in this work coincides with the final status of the very same A and B hydrogenated samples studied in \cite{HYDRO}.

The UHV chamber (base pressure $10^{-9}$ mbar) is equipped with the experimental apparatuses for XPS and EELS. The XPS apparatus consists in an Omicron XM1000 monochromatised Al K$\alpha$ X-ray source (h$\nu$ = 1486.7 eV) and a hemispherical electron analyser (66 mm radius) with position sensitive detector for parallel acquisition. The total energy resolution for this technique is 460 meV and the binding energy scale has been calibrated by setting the C 1s core-level measured for clean highly oriented pyrolytic graphite (HOPG) at 284.5 eV \cite{Chen}. The source of electrons for EELS is a custom-made monochromatic electron gun (45 meV energy resolution) operated at fixed electron energy of 91 eV \cite{Apponi_2021}, the electron analyser is the same used for XPS and the total energy resolution for EELS is 60 meV.

Air exposure and storage were carried out in ambient atmospheric conditions with temperature (24 $\pm$ 1) $^\circ$C and (40 $\pm$ 3)$\%$ humidity.

\begin{figure*}
\centering
\includegraphics[width=\textwidth,trim={4cm 8cm 4cm 7cm},clip]{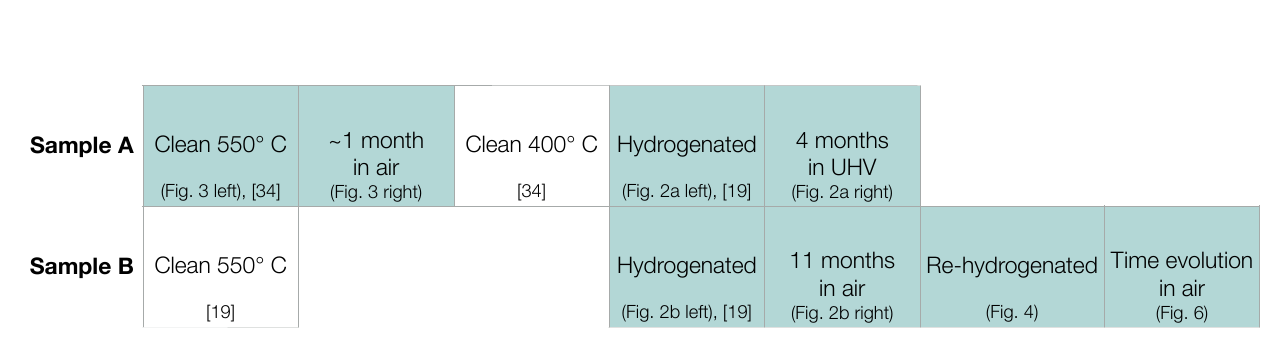} 
\caption{Schematic summary of the experimental steps performed on samples A and B. The shaded blue-grey blocks represent the steps discussed in this work, each with a reference to its corresponding figure. The withe blocks refer to experimental steps not reported in this work, the details of which can be found in \cite{Apponi2024, HYDRO}.}
\label{fig:Figure1}
\end{figure*}

For clarity, the experimental steps relevant to the investigations reported in this work, and thoroughly discussed in Section \ref{sec:ExpRes}, are summarised in Figure \ref{fig:Figure1}. 

\section{Experimental results}
\label{sec:ExpRes}

\subsection*{In-vacuum and in-air stability of hydrogenated graphene}

The stability of hydrogen bonded to monolayer graphene was investigated by measuring the C 1s core-level time evolution: sample A was kept in the UHV chamber and measured after 4 months; sample B was taken out of the UHV chamber and remained in air (ambient atmospheric conditions) for 11 months before being measured again. A fit analysis has been carried out on the C 1s spectra in order to deconvolve the contributions of carbon atoms in different chemical configurations. In particular, a global fitting procedure was employed , in which a set of core level spectra are fitted at one time with two types of free parameters: global and individual. The global free parameters are the same for all the fitted spectra, while the individual ones are specific for a sigle spectrum in the set (more details on this procedure can be found in \cite{PAOLONI23, HYDRO}). This procedure was performed on all the C 1s spectra presented in this work relative to both samples at each experimental step. In the $sp^2$ component, the Doniach-Sunjic line-shape parameters (asymmetry, Gaussian width and Lorentzian width) were free global parameters. The other components ($sp^3$, C-O-C, O-C=O) were fitted with symmetric profiles, where the binding energy shifts, with respect to the $sp^2$ position, and the Gaussian and Lorentzian widths were free global parameters. The areas of the components in all the spectra were free individual parameters instead. 

\begin{figure}
\centering
\includegraphics[width=\columnwidth,trim={5cm 2cm 5cm 2cm},clip]{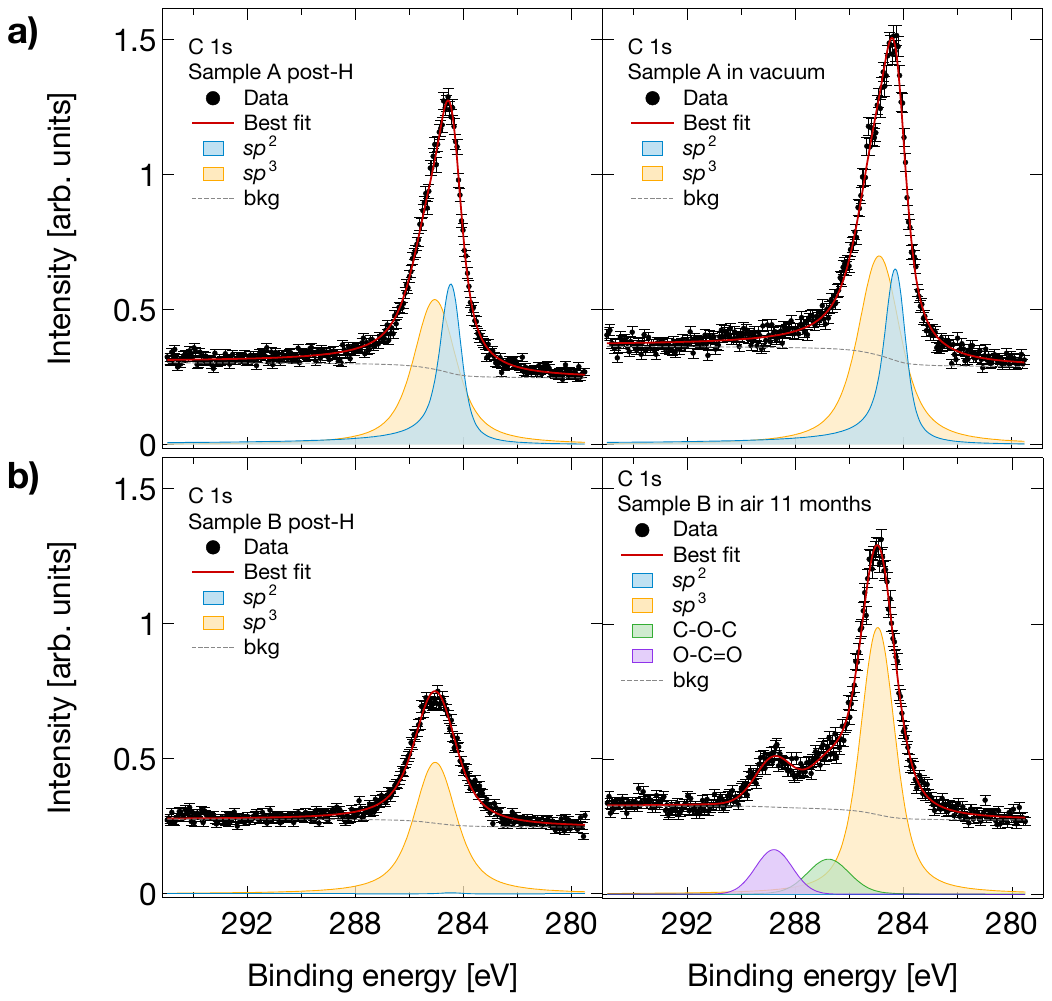} 
\caption{Fit analysis of the C 1s core-level spectra: (a) sample A after hydrogen exposure (left) and after 4 months in UHV (right); (b) sample B after hydrogen exposure (left) and after 11 months in air (right). The black dots represent the experimental data, the best fit curve is shown with red solid line, the shadowed curves are the $sp^2$ (blue), $sp^3$ (yellow), C-O-C (green) and O-C=O (violet) components and finally the dashed grey line is the integral background.}
\label{fig:Figure2}
\end{figure}

In Figure \ref{fig:Figure2}, the measured C 1s spectra along with the results of the global fit analysis for samples A and B, after exposure to hydrogen and after 4 months in vacuum and 11 in air respectively, are shown. For sample A, which was kept in vacuum, a high stability of the hydrogenated monolayer graphene is observed. Indeed, the $sp^3$ relative intensity, represented by the area ratio $\frac{sp^3}{sp^2+sp^3}$, is (61 $\pm$ 2)$\%$ after hydrogenation and (65 $\pm$ 2)$\%$ after 4 months in UHV. Thus within the experimental uncertainty the C 1s line-shape resulted almost unchanged. In the case of sample B, which reached an almost 100$\%$ $sp^3$ saturation after hydrogen exposure, the 11 months in air led instead to a significant oxidation. The global fit revealed, indeed, the rise of two components that can be attributed to carbon bonded to oxygen in different configurations: the component at 286.8 eV associated to C-O-C and the component at 288.8 eV related to O-C=O \cite{Liscio, STOBINSKI, PERROZZI}. Finally, the increase in the intensity of the $sp^3$ component can be attributed to adsorbed carbon-based contaminants on the surface, non resolved in the spectrum.

\begin{figure}
\centering
\includegraphics[width=\columnwidth,trim={5.8cm 5.5cm 5.8cm 6cm},clip]{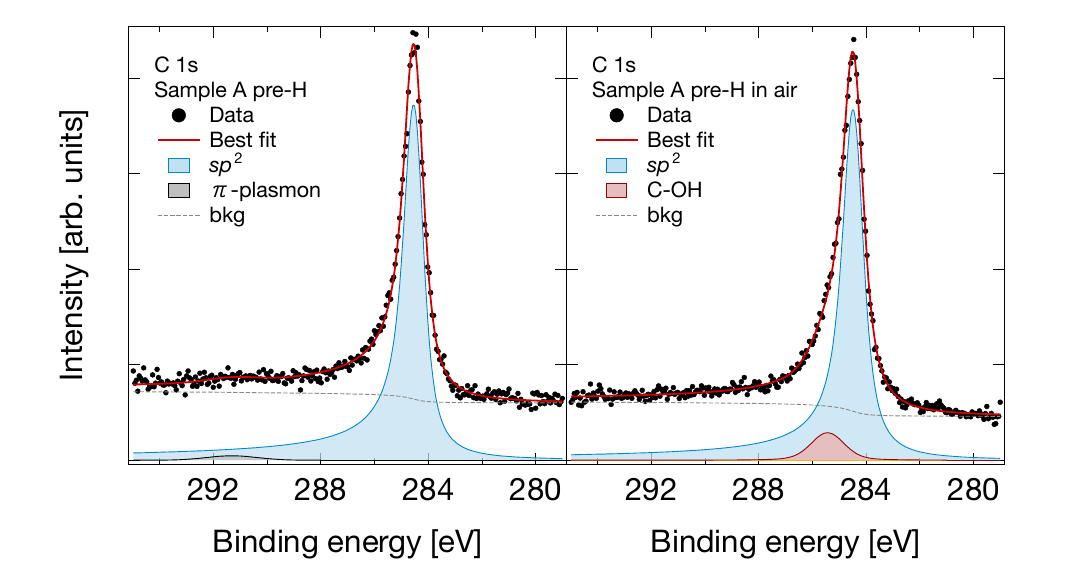} 
\caption{C 1s core level spectrum of non-hydrogenated sample A before and after air-exposure, along with a fit analysis. Representation follows same criteria used in Figure \ref{fig:Figure2} except for $\pi$-plasmon component shown in grey and C-OH component in red.}
\label{fig:Figure3}
\end{figure}

Figure \ref{fig:Figure3} shows the C 1s spectrum of non-hydrogenated sample A before and after air-exposure. The measurement was carried out before any hydrogenation treatment performed on sample A. By exposing clean graphene to air for 28 days, the carbon-oxide C-OH component at 285.5 eV \cite{Liscio} is only 8$\%$ of the C 1s total area. Thus, the non-hydrogenated sample shows a significantly lower oxygen contamination after air exposure with respect to the hydrogenated ones. This result represents a clear indication of hydrogen-induced oxidation of graphene.

\subsection*{Re-hydrogenation of oxidised graphene}

\begin{figure}
\centering
\includegraphics[angle=-90,width=\columnwidth,trim={0cm 2cm 4.5cm 3cm},clip]{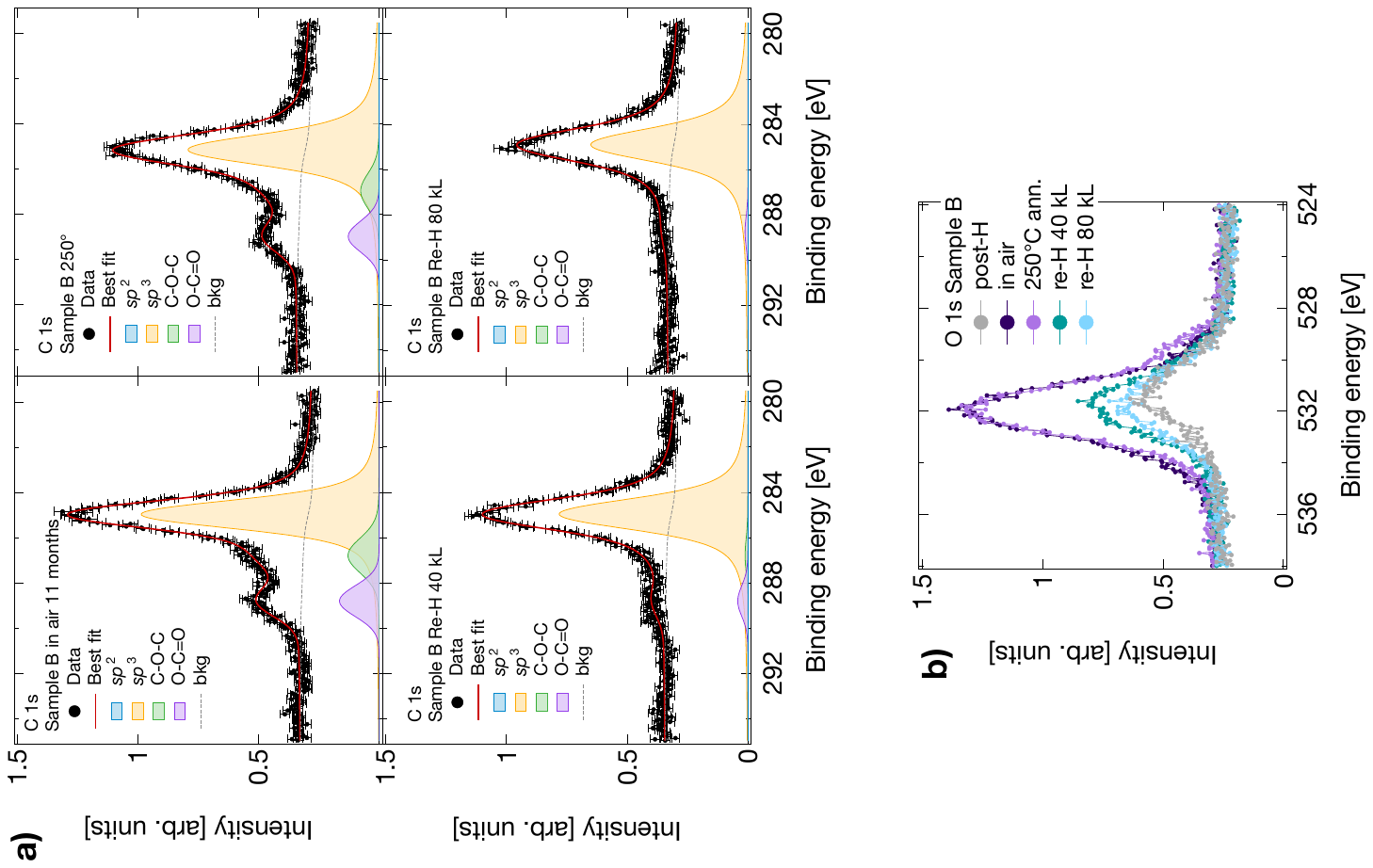} 
\caption{(a) Fit analysis of the C 1s core level for each step: 11 months in air, 250 $^\circ$C annealing, first (40 kL) and second (80 kL) re-hydrogenation. Color coding follows same criteria used in Figure \ref{fig:Figure2}. (b) O 1s core-level spectra of sample B measured after hydrogenation (grey), 11 months in air (purple), 250 $^\circ$C annealing (violet), first (cyan) and second (light blue) step of re-hydrogenation.}
\label{fig:Figure4}
\end{figure}

The \emph{reversibility} of the oxidation process of hydrogenated graphene was studied on sample B. It was firstly annealed at 250 $^\circ$C in vacuum in order to remove adsorbed contaminants and then exposed again to atomic hydrogen in two steps of 40 kL each. The effects on the C 1s line-shape can be observed in Figure \ref{fig:Figure4}(a) with the fit analysis. 

The 250 $^\circ$C annealing of the sample produced a significant reduction of the $sp^3$ intensity, probably due to the desorption of contaminants from the surface, while it has no clear effects on the graphene oxidation (C-O-C and O-C=O components). The latter are, on the contrary, strongly affected by re-exposure of the sample to atomic hydrogen: significantly reduced after 40 kL and almost completely removed with 80 kL. The consequences of these treatments find confirmation in the behaviour of the O 1s core-level spectrum, shown in Figure \ref{fig:Figure4}(b). The annealing does not affect significantly the intensity of the spectrum, while it is drastically reduced after H-exposure. It is important to point out that the measured oxygen can come from both the graphene and the nickel of the TEM grid, bonded or adsorbed. However, the excess in the O 1s spectrum around 532.2 eV of the re-hydrogenated (80 kL) graphene with respect to the hydrogenated one is compatible with the binding energy of O-C=O \cite{STOBINSKI}. This is in agreement with the C 1s fit, in which a residual O-C=O is still present after re-hydrogenation (80 kL). According to the literature, this excess can be also ascribed to residual nickel oxides of the supporting grid \cite{Liu}.

\begin{figure}
\centering
\includegraphics[width=\columnwidth,trim={7.5cm 5.5cm 7.5cm 6cm},clip]{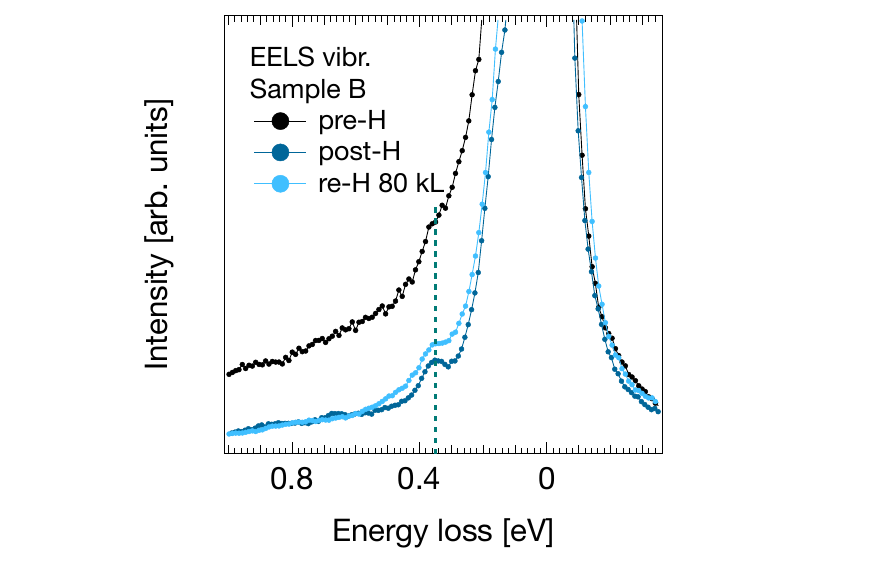} 
\caption{Vibrational EELS spectra measured on sample B before the hydrogenation (black), after the hydrogenation (blue) and after the re-hydrogenation (light blue). The dashed line guides the eye to the peak associated to the CH-stretching at 350 meV.}
\label{fig:Figure5}
\end{figure}

After re-exposure to atomic hydrogen of sample B, the CH-stretching vibrational mode was measured with EELS as a further confirmation, and direct footprint, of re-hydrogenation \cite{DiFilippo}. The spectra were acquired in the vibration region in order to study the evolution of the CH-stretch mode, before and after re-hydrogenation. The EELS spectra for sample B before hydrogen exposure, after first exposure and after re-exposure (80 kL) are shown in Figure \ref{fig:Figure5}. The CH-stretching is well visible at 350 meV after both hydrogenation and re-hydrogenation as a direct evidence of the hydrogen bonding to graphene (further details on the EELS spectrum of sample B after the first hydrogenation can be found in \cite{HYDRO}).

\subsection*{Oxidation time scale of air-exposed hydrogenated graphene}
\begin{figure}
\centering
\includegraphics[angle=-90,width=\columnwidth,trim={0cm 2cm 0cm 3cm},clip]{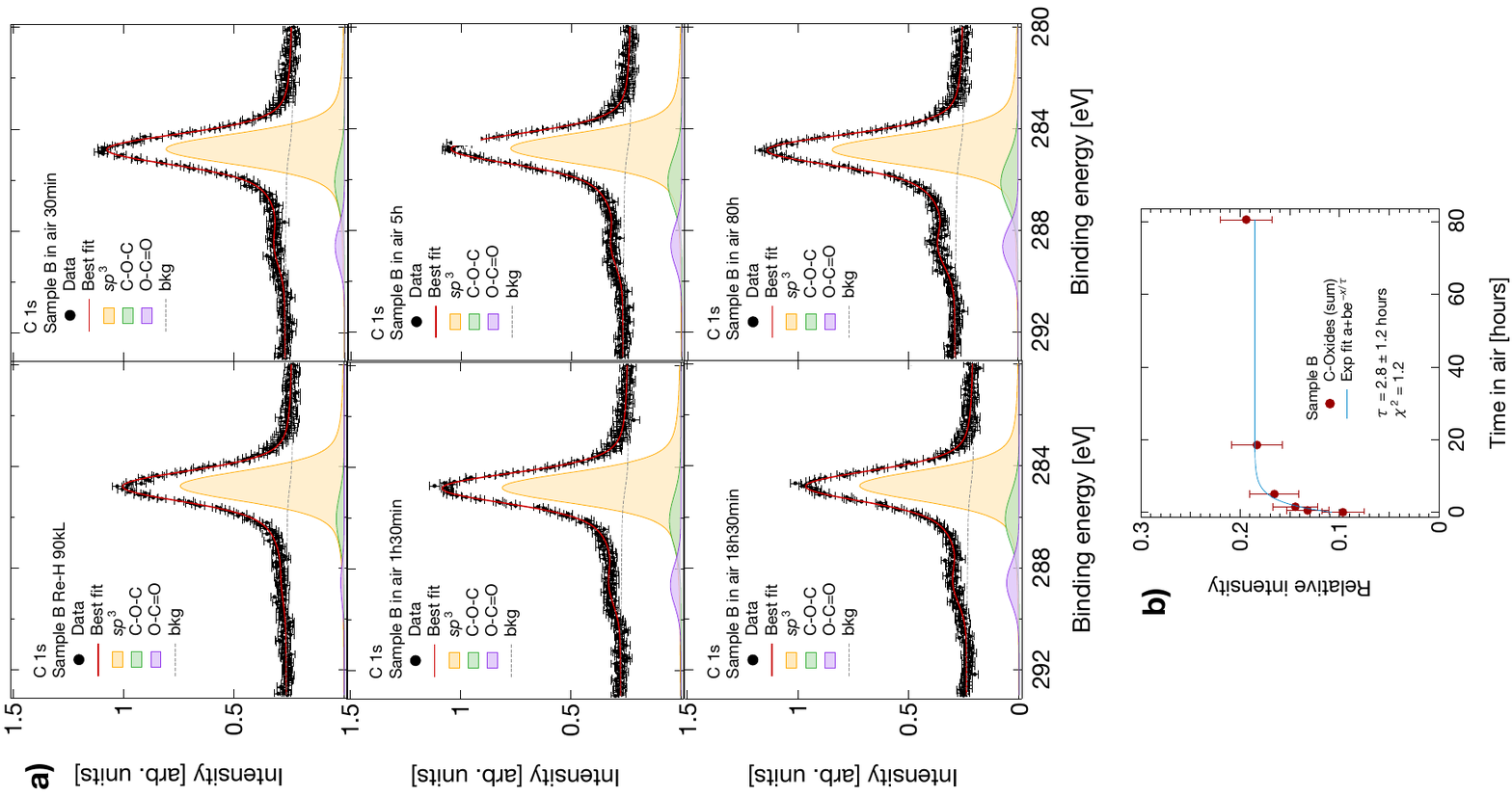} 
\caption{(a) C 1s spectra, with fit analysis, measured on sample B after a 90 kL re-hydrogenation and after 0.5, 1.5, 5, 18.5, 80 hours of (cumulative) air exposure. Curve representation follows same criteria used in Figure \ref{fig:Figure2}. (b) Sum of C-O-C and O-C=O intensities over the total C 1s intensity (red dots) versus air-exposure time of sample B, with exponential fit (blue line). First point at time 0 hours refers to re-hydrogenated sample.}
\label{fig:Figure6}
\end{figure}

With the aim of studying the oxidation process time scale of hydrogenated graphene, the C 1s core-level was measured as a function of time in air. The experiment was performed on sample B after a 90 kL re-hydrogenation and by keeping the sample out of the UHV chamber for 0.5, 1.5, 5, 18.5, 80 hours (cumulative). At each step the C 1s core-level was measured and the spectra are reported in Figure \ref{fig:Figure6}(a), along with the fitting curves. The relative intensity of the carbon oxides (sum of C-O-C and O-C=O components over the total C 1s intensity) follow an exponential behaviour up to saturation as a function of the air-exposure time, as shown in Figure \ref{fig:Figure6}(b). Thus, the time-scale of the oxidation process for hydrogenated graphene has been evaluated through a correlated exponential fit ($\mathrm{a+b}e^{-x/\tau}$), which takes into account the covariance matrix associated to the carbon oxide relative intensities. A time constant $\tau$ = 2.8 $\pm$ 1.2 hours resulted from the fit (reduced $\chi^2$ = 1.2).

\section{Discussion}
\subsection*{Stability of hydrogenated graphene and re-hydrogenation}
The herein reported results show a long-term stability - over four months - of hydrogenated graphene in ultra-high vacuum, in contrast to its strong reactivity under atmospheric conditions. Conversely, clean graphene resulted to be highly stable in air, although not completely inert, thus indicating a hydrogen-induced reactivity of graphene upon air exposure. Despite the significant oxidation in air, the recovery of hydrogenated graphene after re-exposure to atomic hydrogen was achieved. The latter process is far from surprising, as exposure to atomic hydrogen is, among others, a known method for the reduction of graphene oxide \cite{Heya, Jelea}. Atomic hydrogen can indeed react on the graphene surface and remove oxygen, through the formation of hydroxyl radicals OH. Depending on the carbon-oxygen bonding, three possible processes may occur on the surface, although they should be considered speculative in nature \cite{Heya, Jelea}:

\begin{enumerate}
\item{C-O-C + 2H $\rightarrow$ C-OH + C-H;}

\item{C=O + 2H $\rightarrow$ H-C-OH;}

\item{O=C-OH + H $\rightarrow$ HO-C-OH.}
\end{enumerate}

Under further reaction of OH radicals with H atoms, hydrogenated graphene can be restored as follows:
\begin{center}
C-OH + 2H $\rightarrow$ C-H + H$_2$O.
\end{center}

Conversely, the oxidation process of hydrogenated graphene in atmospheric conditions is still not clear. There are a few available studies on the stability of hydrogenated graphene in different environments, where dehydrogenation \cite{Whitener} and oxidation \cite{Kaczmarek} are observed. The investigations reported in \cite{Whitener} show a different reactivity of hydrogenated graphene depending on the substrate, observing a dehydrogenation and recovery of graphene upon exposure to chemical oxidants. On the other hand, an increase in the oxidation potential for hydrogenated graphene is highlighted in \cite{Kaczmarek}, experimentally confirmed by signs of dehydrogenation and oxidation of a graphene sample after exposure to air. The latter result is in better agreement with the observations reported in this work. However, the discrepancy among the three experiments can be reasonably ascribed to the difference in both the investigated graphene samples and, even more, in the oxidation method: with chemical oxidants in \cite{Whitener}, with air exposure in this work and in \cite{Kaczmarek}.

\subsection*{Considerations on the tritium storage with graphene}
In the context of using graphene as a solid-state reservoir for atomic tritium, in a neutrino physics experiment - such as PTOLEMY \cite{Cocco, PTOLEMY4, Apponi_2022} - as well as in fusion energy experiments \cite{Treview}, the results reported in this work provide a clear indication: tritiated (hydrogenated) graphene should be kept in vacuum to preserve the surface functionalisation and, no less important, to avoid the release of tritium in air. However, while for hydrogenated graphene the in-vacuum condition is the only concern for the surface stability, in the case of tritiated graphene the $\beta$-decay could contribute to sample degradation as well. In particular, the emitted $\beta$-electrons and the recoiling $^3$He$^+$ ions may damage the graphene lattice (\emph{radiolysis}). In this regard, some rough considerations can be done to evaluate the relevance of the radiolysis issue. 

Let us firstly focus on the $\beta$-electrons. From theoretical calculations on the $\beta$-spectrum for tritium bonded to single-layer graphene \cite{ACmaster, casale2025}, the rate of emitted electrons as a function of their kinetic energy can be obtained. If two possible scenarios for the final state - $^3$He$^+$ bonded to graphene in the ground state and a free $^3$He$^+$ - are considered, the rate of emitted electrons for 1 $\mu$g of tritium results in a distribution with mean energy of 5.9 keV and end-point at 18.58 keV \cite{biblionote}. The electrons that can cause damage to the surface are the ones with low energy ($\leq$1 keV) \cite{Apponi2024}. The electron current in this energy region is in the order of a few fA/eV, thus the integral of the current up to 1 keV will be of a few pA. This current is a factor 100 lower than the typical values we use for EELS measurements, such as the ones reported in the previous section, in which the sample is irradiated for more than 6 hours and no sign of sample damage has been ever observed. Furthermore, the electron beam employed for spectroscopy is focused on a small portion of the sample ($\sim$0.5 mm$^2$), while for 1$\mu$g of tritium the $\beta$-electrons would be spread on the whole surface of $\sim 10^3$ fully tritiated samples and emitted over the full solid angle (4$\pi$). Therefore, from the comparison with EELS measurements, in which the current density is at least a factor $10^6$ higher, we expect the $\beta$-electrons to be harmless for the graphene surface.

On the other hand, the $^3$He$^+$ ions recoil with kinetic energies of a few eV for each $\beta$-electron emitted up to end-point. Thus, the ion current is at most a few tens of pA and it is spread on the whole surface of $\sim 10^3$ fully tritiated samples and over the full solid angle. An extreme opposite case can be considered for comparison: ion sputtering. Sputtering treatments are usually carried out in solid-state physics experiments to clean surfaces. Damages are intentionally induced by ripping atoms from the sample surface, typically bombarding it with an Ar ion beam for several minutes. The current density is in the order of 100 nA/mm$^2$, the energy of the order of 1 keV. Therefore, even neglecting the ion energy $10^3$ times lower, the current density for the $^3$He$^+$ ions in the $\beta$-decay is at least a factor $10^8$ lower with respect to the typical values used for sputtering.

To conclude, the comparison with EELS and Ar sputtering, one harmless and the second damaging but at least a factor $10^8$ more intense than in our case, allows us to speculatively infer that radiolysis may not represent a critical issue for tritiated graphene. However, specific experiments will have to be carried out to confirm this argument.

\section{Conclusions}
The experimental results reported in this work show a long-term stability of hydrogenated graphene in vacuum and the oxidation of the surface when exposed to atmospheric conditions. The study has been carried out with XPS measurements of the C 1s and O 1s core levels. The hydrogenated graphene sample that was kept in UHV shows a high stability over four months, with a C 1s spectrum almost unchanged within the experimental uncertainty. Conversely, a significant oxidation of the sample exposed to air was observed. In order to determine the time-scale of the in-air oxidation process, the C 1s spectrum of the sample was systematically measured as a function of the air exposure. The carbon oxides relative intensity shows a rapid growth up to saturation with a time constant $\tau$ = 2.8 $\pm$ 1.2 hours. Finally, the recovery of hydrogenated graphene after oxidation was obtained with the re-exposure of the sample to atomic hydrogen.

The sample degradation when exposed to atmospheric conditions results in the partial dehydrogenation and oxidation of the hydrogenated graphene. This outcome finds good agreement with the studies reported in \cite{Kaczmarek}, where signs of C-O bonding formation were observed on a hydrogenated graphene sample after air exposure.

In conclusion, graphene is a good candidate for hydrogen (or tritium) storage as long as it is kept in vacuum, to prevent dehydrogenation. In the case of tritiated graphene, rough considerations on the radiolysis have been pointed out on the basis of experimental experience with electron and ion bombardment of the surfaces. The radiolysis seems to do not represent a critical issue for tritiated graphene, although experimental confirmations of this hypothesis will be required further on.

\section*{Acknowledgements}
We are grateful to Gianfranco Paruzza of the INFN Roma Tre Mechanical Workshop for his technical support. We gratefully acknowledge financial support from MUR PRIN 2020 'ANDROMeDa' project (Contract No. PRIN 2020Y2JMP5). This project has received funding from the European Union's Horizon 2020 research and innovation programme Graphene Flagship under grant agreement No. 881603. We are also grateful to INFN CSN5 for the financial support in this research.

\bibliography{bibliographyGrH_stability}


\end{document}